\documentclass[ 
reprint,
 amsmath,amssymb,
 aps,showpacs,superscriptaddress
]{revtex4-1}

\usepackage{graphicx}
\usepackage{dcolumn}
\usepackage{bm}
\usepackage{subfig}
\usepackage{cancel}
\usepackage{mathtools} 
\usepackage{upgreek}

\usepackage{color}
\usepackage{xcolor}

\usepackage{enumitem}
\newcommand{\beqar}{\begin{eqnarray}}
\newcommand{\eeqar}{\end{eqnarray}}

\newcommand{\bcen}{\begin{center}}
\newcommand{\ecen}{\end{center}}

\newcommand{\eps}{\varepsilon}

\newcommand{\f}[2]{\frac{#1}{#2}}
\renewcommand{\b}[1]{\left({#1}\right)}
\renewcommand{\v}[1]{\vec{#1}}

\renewcommand{\sb}[1]{\left[{#1}\right]}

\newcommand{\ra}{\rightarrow}

\begin{document}

\preprint{APS/123-QED}
\title{Variational solutions for Resonances by a Finite-Difference Grid Method}

\author{Roie Dann}
\thanks{These two authors contributed equally}
\email{roie.dann@mail.huji.ac.il}
\affiliation{The Institute of Chemistry, The Hebrew University of Jerusalem, Jerusalem 9190401, Israel.}
\author{Guy Elbaz}%
\thanks{These two authors contributed equally}
\email{guyelbaz3@gmail.com}
\affiliation{Faculty of Mechanical Engineering, Technion–Israel Institute of Technology, Haifa 3200003, Israel}%
\author{Jonathan Berkheim}%
\affiliation{School of Chemistry, Tel Aviv University, Tel Aviv, 69978, Israel}%
\author{Alan Muhafra}%
\affiliation{Faculty of Mechanical Engineering, Technion–Israel Institute of Technology, Haifa 3200003, Israel}%
\author{Omri Nitecki}%
\affiliation{Schulich Faculty of Chemistry, Technion-Israel Institute of Technology, Haifa 3200003, Israel}%
\author{Daniel Wilczynski}%
\affiliation{School of Chemistry, Tel Aviv University, Tel Aviv, 69978, Israel}%
\author{ Nimrod Moiseyev }%
\email{nimrod@technion.ac.il}
\affiliation{Schulich Faculty of Chemistry, Solid State Institute and Faculty of Physics, Technion-Israel Institute of Technology, Haifa 3200003, Israel}%

\begin{abstract}
    We demonstrate that the finite difference grid method (FDM) can be simply modified to satisfy the variational principle and enable calculations of both real and complex poles of the scattering matrix. These complex poles are known as resonances and provide the energies and inverse lifetimes of the system under study (e.g., molecules) in metastable states.  This approach allows incorporating finite grid methods in the study of resonance phenomena in chemistry.  Possible applications include the calculation of electronic autoionization resonances which occur when ionization takes place as the bond lengths of the molecule are varied. Alternatively, the method can be applied to calculate nuclear predissociation resonances which are associated with activated complexes with finite lifetimes.
\end{abstract}
\maketitle
\section{Introduction}

In 1978 Frank Weinhold together with Phil Certain and Nimrod Moiseyev,  in their work on the complex variational principle (a stationary point rather than an upper bound as in Hermitian QM), paved the way for the use of electronic structure computational algorithms to metastable (resonance) states \cite{moiseyev1978resonance}. In this framework, the energies and inverse lifetimes of atoms and molecules are associated with the real and imaginary parts of the complex eigenvalues of  non-Hermitian Hamiltonians, respectively. 


Recently, there is increasing interest among chemists concerning the use of non-Hermitian QM for the calculations of molecular resonances. For example, the most recent developments of the QCHEM quantum chemistry package enable calculations of shape type and Feshbach molecular resonances \cite{bravaya2013complex,jagau2014complex,zuev2014complex,jagau2014fresh,jagau2015same,kunitsa2015first,jagau2016investigating,jagau2016characterizing,kunitsa2017cap,jagau2017extending,benda2017communication,benda2018locating,jagau2018non,li2019dipole,hernandez2019resolution,hernandez2020resolution,parravicini2021embedded}. In addition, such methods have been employed in the study of RNA bases \cite{fennimore2016core,fennimore2018electronic}. It is clear by now that in order to increase the accuracy of ab-initio calculations, associated with resonances,  grid methods should be incorporated into the Gaussian basis set. See for example the hybrid Gaussian and b-spline method recently developed by Fernando Martin and his group \cite{marante2014hybrid,marante2015merging}. Specifically, this will enable to improve the description of the outgoing ionized electrons.
However, the standard finite difference method (FDM), that enables calculation of the bound states of a quantum system, is not applicable for the calculations of the complex poles associated with metastable (resonance) states.  The origin of this failure can be traced back to the fact that the standard FDM does not satisfy the variational principle.

Here we  show how a simple change in  the  selection  of  the  grid points in FDM leads to a variational principle and enables calculation of both real and complex poles of the scattering matrix.  This approach opens the gate to evaluate the resonances by FDM for atoms and molecules as well as mesoscopic systems.  Illustrative numerical examples will be given for a “toy” model{,} which was first presented in a paper published together with Frank Weinhold many years ago and is used until now as a model for testing new algorithms for calculating resonances \cite{moiseyev1978resonance}.

Numerical approaches for the analysis of physical systems can be classified into two prominent categories: grid and basis set approaches \cite{ditchfield1971self,davidson1986basis,lewars2003computational,hinchliffe1996modelling,szabo2012modern,levine2009quantum,huey2004grid}. The basis set methods are equivalent to the use of an approximate representation of the identity operator. As a result, they provide upper bounds to the eigenvalues. {Contrastly, }in the grid based methods one represents the continuous space by a quantized finite number of grid points. These methods exhibit fast processing time, however, they generally do {not}  provide an upper bound to the eigenvalues \cite{forsythe1960finite,perrone1975general,liszka1980finite,chelikowsky1994finite}. 

The standard grid method is  
the traditional finite difference method, which is abundantly used in the solution of second order partial {differential equations, e.g., in the study of} heat transfer problems \cite{ozicsik2017finite}, {as well as} in solving the Maxwell \cite{teixeira2008time,holland1983finite} and Schr\"odinger equations \cite{simos1997finite}. The crucial limitation of the standard FDM is that the convergence of the numerical results requires refining the grid spacing (mesh), which
in turn increases the amount of storage and calculation. An important improvement of  the accuracy and stability of the FDM has been recently described in Ref. \cite{chen2020new} by combining two high-order exponential time differencing precise integration methods (PIMs)  with a spatially global sixth-order compact finite
difference scheme (CFDS).  In addition, by modifying the {representation of the Laplacian operator} one can obtain a rigorous upper bound estimate of the true kinetic energy  \cite{maragakis2001variational}.

The first goal of this paper is to show that upper bounds to the spectrum of any given Hamiltonian can be obtained without modifying the {representaion of the Laplacian} and by using the same set of coupled equations as are used in the standard FDM. We refer to the proposed method as the ``{\emph{present}}'' variational FDM, while the common approach is termed the ``{\emph{standard}}'' non-variational method.
The standard FDM typically converges to the exact spectrum from {below} (i.e., non-variational), this is attributed to the fact that the obtained spectrum of the kinetic energy operator in the standard FDM serves as a lower bound to the exact kinetic energies, see Fig. \ref{fig:KE_N_m1} and also Ref. \cite{maragakis2001variational}. Note however that this characteristic behaviour is not true for any potential.

 Building upon the Hylleraas Undheim and MacDonald theorem  we prove that the proposed present FDM satisfies a variational principle with respect to the accurate solution within the finite box approximation. The variational principle guarantees the  stability of the proposed scheme as the number of the grid points are increased. The stability of the present FDM calculations is obtained by holding the grid spacing to be as small as possible and constant, while increasing the number of grid points.

We first focus on the calculation of the bound discrete states of Hermitian Hamiltonians. Following, we show how the the present FDM can be utilized to calculate the energies and widths (inverse lifetimes) of mestasbale states states, embedded in the continuous part of the spectrum (so called resonances), which are associated with the poles of the scattering matrix.

We introduce the finite box quantization condition, assuming that this restriction does not serve as a limitation to obtain the bound state spectrum in the desired accuracy.
That is, the exact result is considered as the result obtained by fixing the spatial range of the system and infinitely increasing the precision of the calculation. 
Physically, this is motivated by the fact that any realistic computation is conducted by using {a} finite number of grid {points} or basis states, i.e., finite size computers. Moreover, any realistic measurement has a corresponding fundamental uncertainty.  Therefore, one can replace infinite space by a box of finite dimension, without practically effecting the physical description.

Within the finite-box approximation, the considered exact solution is determined by  two parameters: the box size $L_\text{max}$ and the maximum number of grid points $N_\text{max}$.
In the present approach the grid spacing is defined by $\delta x=L_\text{max}/N_{\text{max}}$, where the box-size $L$ is varied with the number of grid points $N$, i.e., for $L=\delta x N$. This should be distinguished from the standard method, where the grid spacing $\Delta x = L_\text{max}/N$ varies with the number of grid points where $N\leq N_\text{max}$. 

The paper is organized as follows. First, we describe the two FDMs procedures, leading to the spectrum of the Hamiltonian under study. We then provide a proof that the proposed FDM produces an upper bound to the spectrum of the Hamiltonian within the box quantization condition. {Following}, we present  the numerical results for the calculation of the bound and metastable states (resonances), and compare to the exact results. Finally, we  conclude by emphasizing the generality of our approach.
\section{Methodology}

{}
When conducting a numerical calculation utilizing a grid based method, the Hamiltonian operator $\hat{H}$ is represented by a $N\times N$ dimensional matrix,
\begin{equation}
 H=T+V~~,
 \label{eq:Hamil}
\end{equation} 
where $N$ is the number of grid points and $T$ and $V$ are matrix representations of the kinetic and potential energies operators. The matrix $T$ is calculated by utilizing {$m$  ($m=2l+1$; $l=1,2...$)} grid points to evaluate the Laplacian (second-order derivative). {Specifically, we employ the central finite grid method, where $m$ values of the wavefunction around the grid point $x_i$ (identical number on both sides) are used to approximate the second order derivative of the wave function  (cf. Appendix \ref{appendix} for further details)}.
This leads to the following relation

\begin{equation}
     \left(\Delta x\right)^{2} \left.\frac{d^{2}\psi}{dx^{2}}\right|_{x_{i}}=\sum_{j=-l}^{l}w_j\psi_{i+j}~~,
    \label{eq:weights}
\end{equation}
where $w_j={A}_{3,j+l+1}^{-1}$, and  $A$ is a $m\times m$ matrix, with elements $A_{j+l+1,k+1}={j^{k}}/{k!}$ {and} $j\in \sb{-l,l}$.
This relation determines the matrix elements of the $\b{2l+1}$-diagonal matrix $T$ whose elements are given by
\begin{equation}
 T_{i,i+j}=-\f{\hbar^2 {w_j}}{2\mu \b{\Delta x}^2}~~\text{with}~~j\in \sb{-l,l}, 
\end{equation}
while zero otherwise and $i=1,2,\dots,N$. {Here $\hbar$ is Planck's constant and $\mu$ is the particle's mass.}
Table \ref{tab:weights} gives the coefficients for different values of $m$. We emphasize that the coefficients are symmetric $w_{j}=w_{-j}$; this entails that the kinetic energy matrix is symmetric and real, which in turn implies that it is positive definite.

 \begin{table}[h!]
     \centering
     \begin{tabular}{||c | c c c c c c||} 
         \hline
         $w_{j}$ & $w_{0}$ & $w_{1}$ & $w_{2}$ & $w_{3}$ & $w_{4}$ & $w_{5}$ \\ [0.5 ex] 
         \hline \hline
         $m=3$ & $-2$ & $1$ &  &  &  &   \\ [0.5 ex]  
         \hline
         $m=5$ & $-\frac{5}{2}$ & $\frac{4}{3}$ & $-\frac{1}{12}$ &  &  & \\ [0.5 ex] 
         \hline
         $m=7$ & $-\frac{49}{18}$ & $\frac{3}{2}$ & $-\frac{3}{20}$ & $\frac{1}{90}$ &  & \\ [0.5 ex]
         \hline
         $m=9$ & $-\frac{205}{72}$ & $\frac{8}{5}$ & $-\frac{1}{5}$ & $\frac{8}{315}$ & $-\frac{1}{560}$ & \\ [0.5 ex]
         \hline
         $m=11$ & $-\frac{5269}{1800}$ & $\frac{5}{3}$ & $-\frac{5}{12}$ & $\frac{5}{126}$ & $-\frac{5}{1008}$ & $-\frac{1}{3150}$ \\ [0.5 ex]
         \hline
    \end{tabular}
     \caption{The weights $w_{j}$ in \eqref{eq:weights} for different values of $m$. Here we present only $w_{j}$ for $j\geq 0$ as the weights remain symmetric for every $j$, e.g. $w_{j}=w_{-j}$.}
     \label{tab:weights}
 \end{table}
 
 The number of grid points included in the calculation of $T$ ($m$) has a significant effect on the eigenvalues of the matrix. {This can be witnessed by analyzing the convergence to the exact solution with increasing $n$, see Appendix \ref{apsec:plot} for a graphical representation.}

 \begin{figure}[htb!]
    \centering
    \includegraphics[width=8cm]{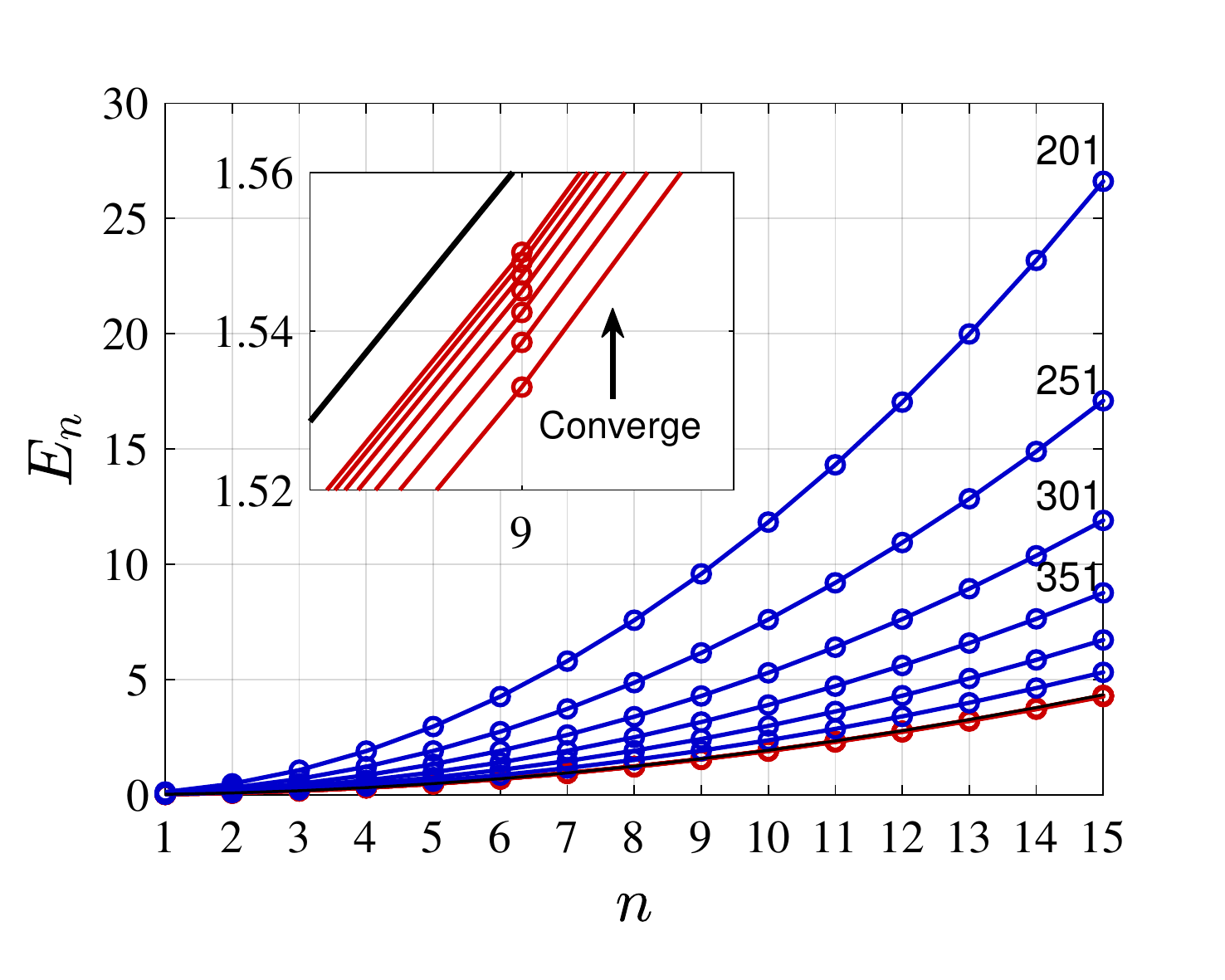}
    \caption{Eigenvalues of the kinetic energy operator $\{E_n\}_{n=1,2,..}$ as obtained by the standard FDM and by the present {(variational)} {FDM}. Red curves are obtained by the {standard} FDM for fixed value of  $L=16$ for growing number of grid points $N$ such that $\Delta X=16/N$.  The blue curves are obtained by the present (variational) FDM for a fixed small grid spacing $\delta x$ and increasing number of grid points. The kinetic energy matrix is calculated by using $m=3$ (see Appendix \ref{appendix}) and therefore is a tri-diagonal matrix. It is evident that the standard method provides lower bounds for the kinetic energy spectrum for every $n\geq 0$ as N is increased (N=201, 251,..), whereas the present (variational) FDM, first presented here, provides upper bounds to the exact values. In this case, the standard FDM converges faster than the present FDM but it is not a variational method.}
    \label{fig:KE_N_m1}
\end{figure}

Alternatively, the {present (variational)} FDM can also provide an upper bound to the spectrum of the Hamiltonian. For a properly defined grid, when the grid-difference is held fixed $\delta x\equiv\Delta x\b{N_{\text{max}}}${, and} the box-size is increased with $N$: $L\b{N} = \delta x N$, the numerical result converges to the exact spectrum from above.  In this case, as seen in Fig \ref{fig:KE_N_m1}, the eigenvalues of  $T$ provide a curve which approaches the parabolic function $y(n)=c(n/L_{\text{max}})^2$, with $L_{\text{max}}=L\b{N_{\text{max}}}$, from above.  As  $N$ approaches the maximal value $N_{\text{max}}$, the obtained spectrum approaches the same converged numerical result of the kinetic energy, leading to the converged Hamiltonian spectrum.

 
\section{The variational principle for finite difference method}
 \label{sec: variation}
We consider a system confined in a finite box {which is} represented by Hamiltonian $\hat{H}$. 
The box size, utilized in the calculation, is chosen so {to} not {limit the accuracy of the} calculated eigenstates.
This is a common approximation, which is implicitly included in any numerical calculation, including all quantum chemistry {packages used to obtain the} electronic spectrum. In such calculations, the molecular Hamiltonian $\hat{H}$ is replaced by a finite dimensional matrix ${H}$. Similarly, in the FDM we limit the 1D space to $a<x<b$. The exact Hamiltonian under study within the box quantization framework is obtained by the FDM when $\Delta x=\b{b-a}/N$ as $N\ra \infty$. 

Based on the Hylleraas, Undheim and MacDonald theorem \cite{hylleraas1930numerische,macdonald1933successive,epstein2012variation}, we prove that the eigenvalues of the $N$-grid point representation matrix of the Hamiltonian, ${H}\b{N}$,  serves as an upper bound to the exact spectrum. The $N$ by $N$ matrix $H\b{N}$ satisfies the following eigenvalue equation
\begin{equation}
    H\b{N}C\b{N} = C\b{N}E^{diag}\b{N}~~,
\end{equation}
where the columns of $C\b N$ are the eigenstates of $H\b{N}$ and $E^{diag}$ is a diagonal matrix containing the corresponding eigenvalues. Clearly, the $N+1$ by $N+1$ matrix $H\b{N+1}$ satisfies a similar eigenvalue equation. This matrix can be expressed in terms of $H\b N$ matrix as follows
\begin{equation}
    H\b{N+1}=\sb{\begin{array}{cc}
H\b N & \v M\\
\v M^\dagger & H_{N+1,N+1}
\end{array}}~~,
\end{equation}
where $\v M = \b{H_{1,N+1},\dots,H_{N,N+1}}^T$ with $H_{ij}$ are the corresponding matrix elements of $H\b{N+1}$. Note that in the present case, where the system is {represented by a grid, $\v M$ never vanishes.}


After some algebraic manipulations one obtains the relation 
\begin{multline}
    \eps\b{N+1}-H_{N+1,N+1}= \v{M}^T C\b{N}\sb{\eps\b{N+1}I-E^{diag}\b{N}}^{-1} C^\dagger\b{N}\v{M}^*
    \label{eq:relation}
\end{multline}  
where $\eps\b{N+1}$ is one of the eigenvalues of $H\b{N+1}$.
Equation \eqref{eq:relation} can be {now} solved by replacing 
$\eps \b{N+1}$ by a parameter $x$ and plotting both sides of the equation as a function of $x$. The intersection between the two curves are values of $x=\eps \b{N+1}$ for which Eq. \eqref{eq:relation} is satisfied.
 \begin{figure}[htb!]
    \centering
    \includegraphics[width=8cm]{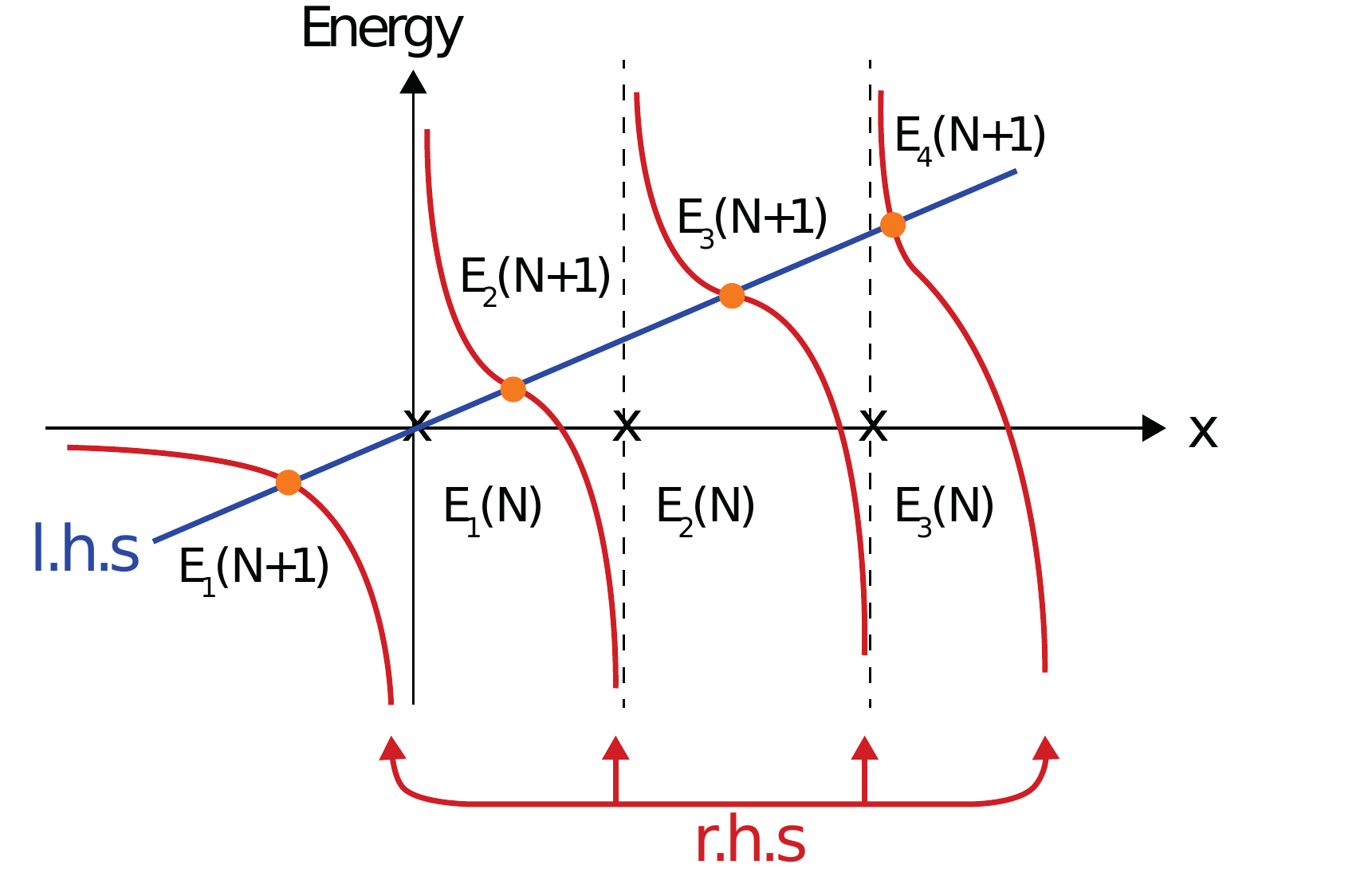}
    \caption{A schematic representation of the both sides of Eq. \eqref{eq:relation} as a function of $x$. Singularities are obtained when $x$ equals one of the eigenvalues of $H\b N$ (crosses on the $x$ axis), while the intersection are achieved when $x$ corresponds to one of the eigenvalues of $H\b{N+1}$ (orange points).}
    \label{fig:scheme}
\end{figure}
Poles of Eq. \eqref{eq:relation} are obtained whenever $\eps\b{N+1}$ is one of the eigenvalues of $H\b{N}$, i.e., one of the elements on the diagonal of $E^{diag}\b{N}$. In Fig. \ref{fig:scheme} a schematic representation of the left hand side (l.h.s) and r.h.s of equation \eqref{eq:relation} are plotted as function of the parameter $x$.  
By observing Fig. \ref{fig:scheme} it is evident that  
\begin{equation}
    E_{{n}-1}\b N<E_{{n}}\b{N+1}<E_{{n}}\b{N}~~.
    \label{eq:10}
\end{equation}
This equation shows that the eigenvalues converge from above. Hence, the eigenvalues obtained using a finite number of grid points upper-bound the exact eigenvalues.  The strict inequality between the eigenvalues of Eq. \eqref{eq:10} emerges from the fact that when the vector $\v{M}$ does not vanish {and the eigenvalue of a matrix with $N+1$ dimensions coincides with  an eigenvalue of a matrix with $N$ dimension, there is a singularity.}
This result completes the proof showing that the present FDM produces an upper bound for the exact solution, {in the desired chosen accuracy, i.e., the exact} solution within the finite box approximation. Upper bounds to the eigenvalues of the Hamiltonian beyond the finite box approximation are obtained for $\delta x\ra 0$.


To demonstrate how the two FDM schemes can be combined together to evaluate the system spectrum,  we compare the FDM results to the analytical solution for two cases: the harmonic and Rosen-Morse potentials.
The harmonic potential, $V_{HO}\b{x}=\f{1}{2} \mu \omega^2 x^2$, includes an infinite number of bounded states with energies $E_n=\hbar\omega\b{n+\f{1}{2}}$, where $n=0,1,2,...$, $\mu$ is the particle mass and $\omega$ is the oscillator frequency. In contrast, the {Rosen-Morse} has finite number of bounded states $n_{\text{max}}$ with energies $$E_n =-\f{\hbar^2a^2}{2\mu}\sb{-\b{1+2n}+\sqrt{1+\f{8\mu V_0}{a^2\hbar^2}}}~~,$$ {where} $n\leq n_{\text{max}}$, and the potential is of the form $V_{RM}=-V_0/\text{cosh}^2\b{a x}$ \cite{rosen1932vibrations}.

The calculation is performed by the following procedure: First, we evaluate the maximum box size $L_{\text{max}}$ utilizing a semi-classical approximation. The semi-classical bound state function is well described when the box quantization condition is imposed on the quantum solution, such that
$|A\b{x=L_\text{max}}|=|\exp\b{-\int\sqrt{2\mu \b{V\b{x}-E_{\text{max}}}}dx}|\approx 0$, where the eigenenergies of interest lie in the range  $\sb{\text{\text{min}}_x\b{V\b{x}},E_{\text{max}}}$.
This evaluation is equivalent to employing the WKB method in order to recast the wavefunction in an exponential form \cite{wentzel1926verallgemeinerung,kramers1926wellenmechanik,brillouin1926mecanique}. {Technically, } this approximation is valid {only} for large action relative to $\hbar$ and smooth potentials, nevertheless, {this condition is sufficient to evaluate  $L_{\text{max}}$ even beyond this regime}. In our calculations we take $A\approx10^{-7}$. For higher dimensional space the box-size should be evaluated in a similar way, by choosing the spatial coordinates according to the classical turning points of the potential in the energy range under study.

\begin{figure}[htb!]
    \centering
    \includegraphics[width=8cm]{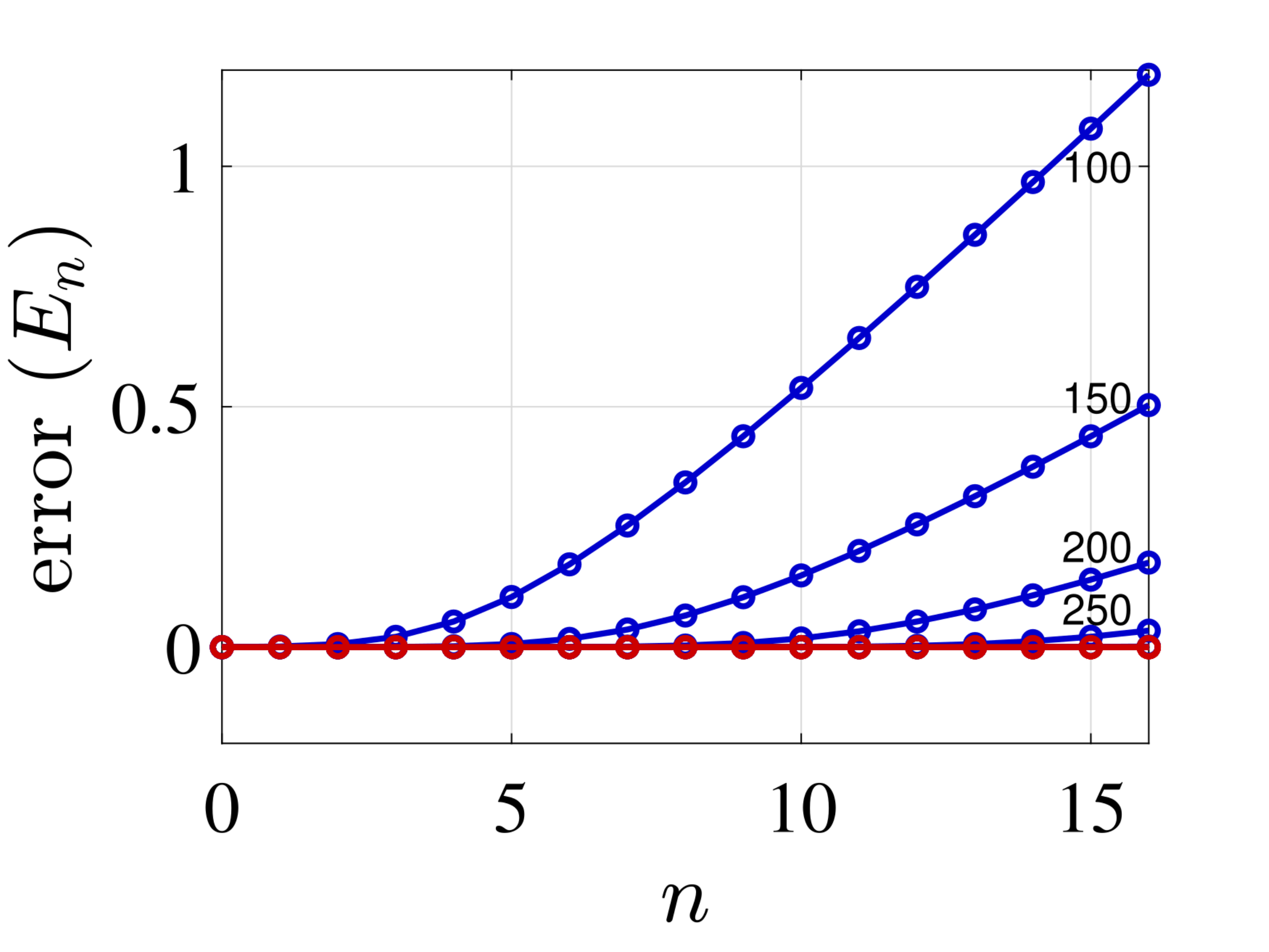}
    \caption{Error of the energy eigenvalues, $n=0,1,2,..$, for the harmonic potential using the standard (non variational) FDM (red curves) and the present (variational) FDM (blue curves). Red curves for fixed value of  $L_{\text{max}}=36$ for growing number of grid points $N$ (N=100,150,..,250), and blue curves $\Delta x=\delta x = L_\text{max}/N_{\text{max}}$. The parameters values are: $\omega=\mu=\hbar=1$ and the kinetic energy is evaluated utilizing  $m=7$ grid points. For the harmonic potential, the upper variational solutions obtained by the present {FDM converge} much {slower relative to}  the non-variational solutions, obtained by the standard FDM. On the scale of this graph it is not possible to follow the convergence of the non-variation (red) results obtained for $N\ge 100$. }
    \label{fig:HO}
\end{figure}

\begin{figure}[htb!]
    \centering
    \includegraphics[width=8cm]{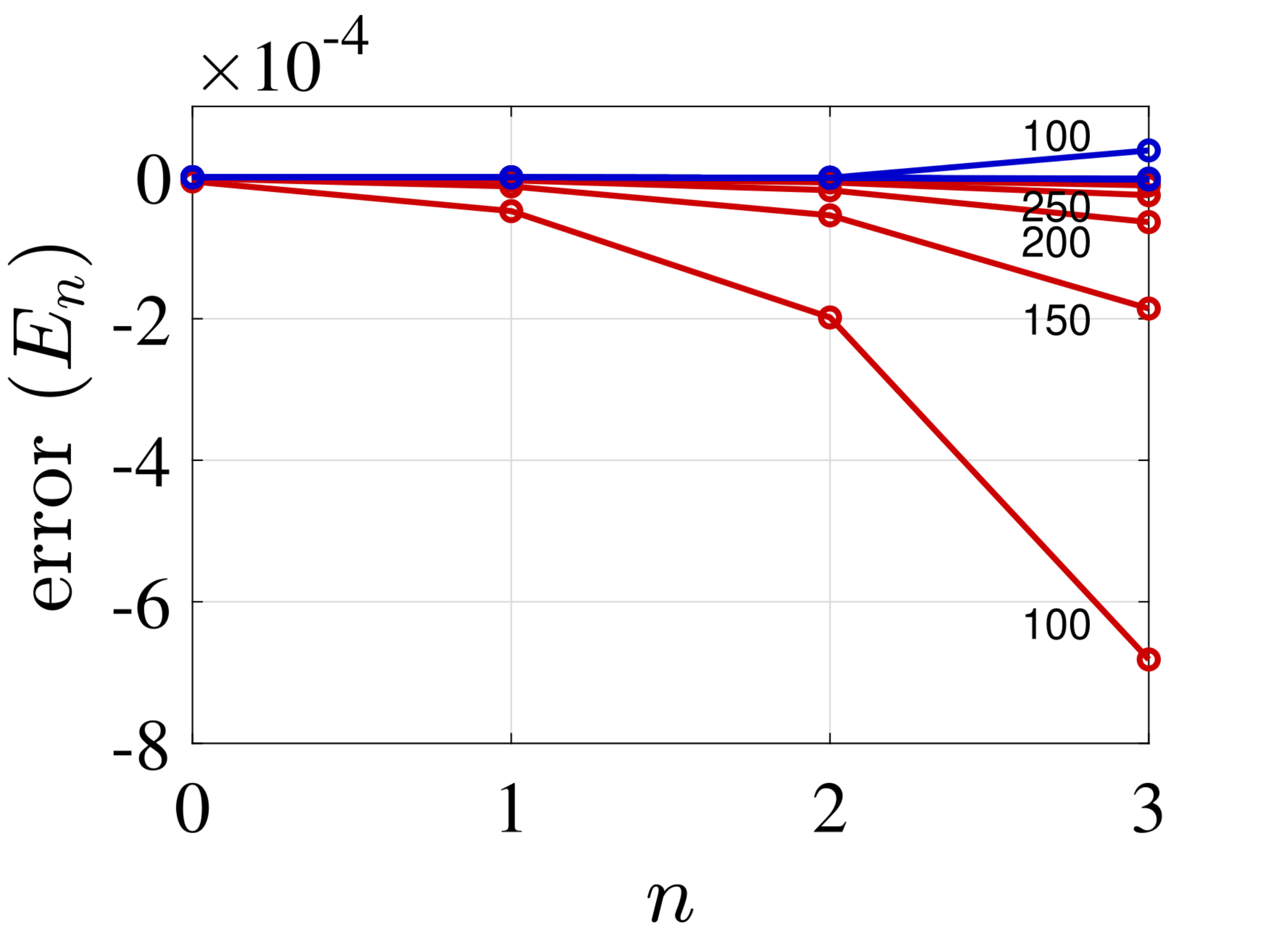}
    \caption{Error of the energy eigenvalues, $n=0,1,2,3$ for the Rosen-Morse potential for different number of grid points $N$ using the standard (non variational) FDM (red curves) and the present (variational) FDM (blue curves). Red curves for fixed value of  $L_{\text{max}}=36$ {with} growing number of grid points $N$ (from N=100 to N=250) where $\Delta x (N)= L_{\text{max}}/N $, and blue curves $\Delta x = \delta x=L_{\text{max}}/N_{\text{max}}$. The parameters values are: $V_0=10$, $a=\mu=\hbar=1$ and the kinetic energy is evaluated utilizing  $m=7$ grid points (i.e., 7th diagonal matrix). As wittnessed in the plot, for this unharmonic potential the upper variational solutions obtained by the present {FDM converge} much faster than the non-variational solutions {which where} obtained by the standard FDM. On the scale of this graph it is not possible to fully follow the convergence of the variation (blue) results obtained for {$N\ge 150$.}}
    \label{fig:rosen_morse}
\end{figure}

{We now compare the eigenenergies} of the two FDM schemes for a varying number of grid points $N$. The standard procedure ($L=\text{const}$), typically, produces a lower bound, while keeping the grid density constant with increasing grid size gives an upper bound to the spectrum. This can be observed in  Figs. \ref{fig:HO} and \ref{fig:rosen_morse} , which  present the energy error, $\text{error}\b{E_n}=\b{E_n^{\text{numerical}}-E_n^{\text{exact}}}/|E_n^{\text{exact}}|$, as a function of quantum number $n$ for the two potentials. The two cases demonstrate the varying convergence behaviour. In the case of the harmonic potential, the standard FDM shows a faster convergence from below relative to the present method. In contrast, the later method shows a rapid convergence for the  Rosen-Morse potential from above. This demonstrates the utility of applying both methods, and combining them to evaluate the exact spectrum. 

\section{Illustrative numerical examples for the calculations of energies and widths of resonances}

We now apply the present FDM to calculate the spectrum of a  model potential 
\begin{equation}
    V_r\b x=\b{x^2/2-0.8}\exp\b{-0.1x^2}~~.
    \label{eq:V_r}
\end{equation}
Such a potential has been used to study new computational algorithms for calculating the energies and widths (inverse lifetimes) of shape type resonances \cite{moiseyev1978resonance}. The spectrum is characterized by a single bound state and metastable states with higher energies. 
In addition, in Ref. \cite{moiseyev1998quantum}
 this model was employed to calculate upper and lower bounds to the complex decay poles of the scattering matrix (resonances). 

\begin{figure}[htb!]
    \centering
    \includegraphics[width=6.6cm]{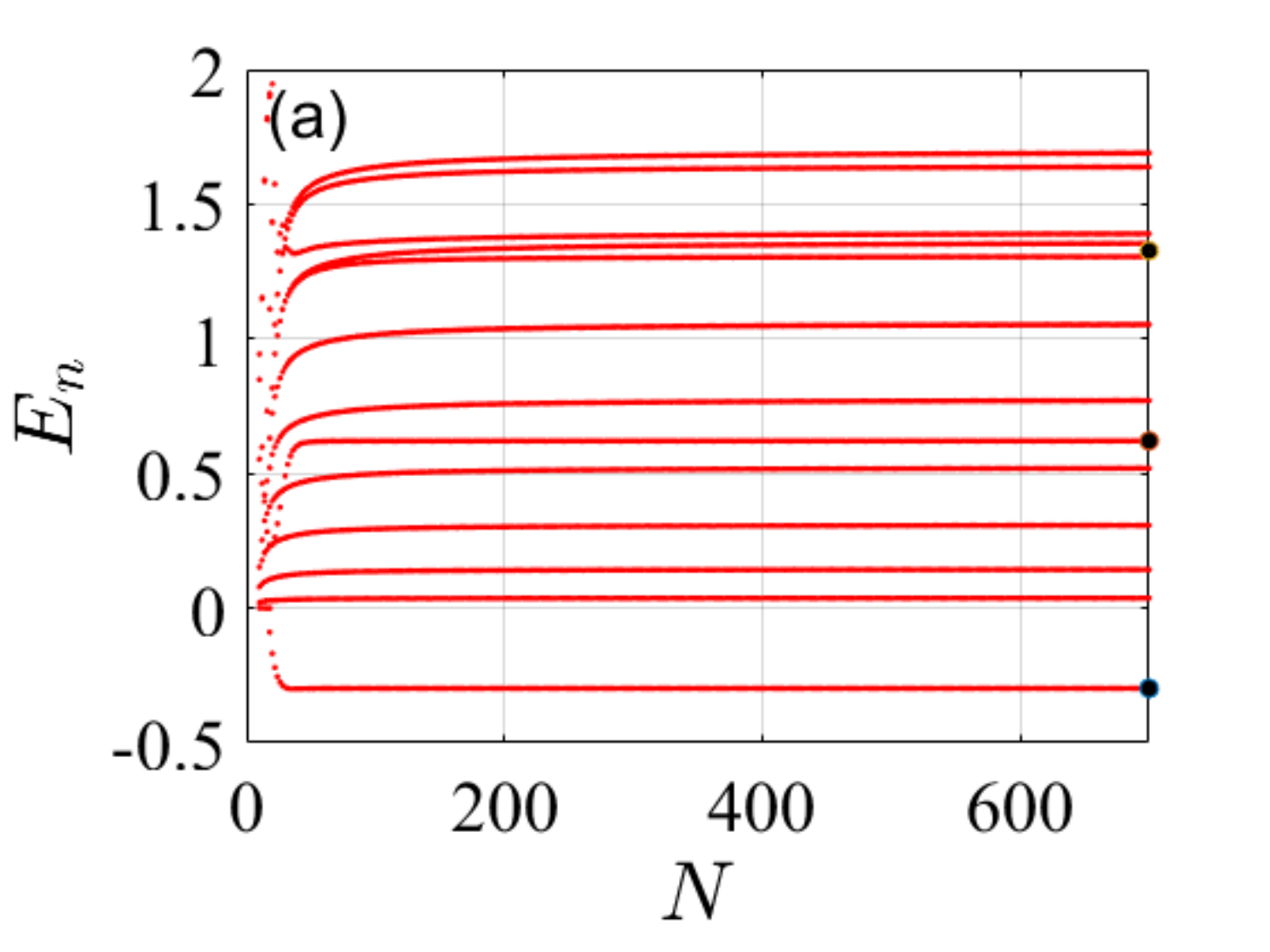}
    \includegraphics[width=6.6cm]{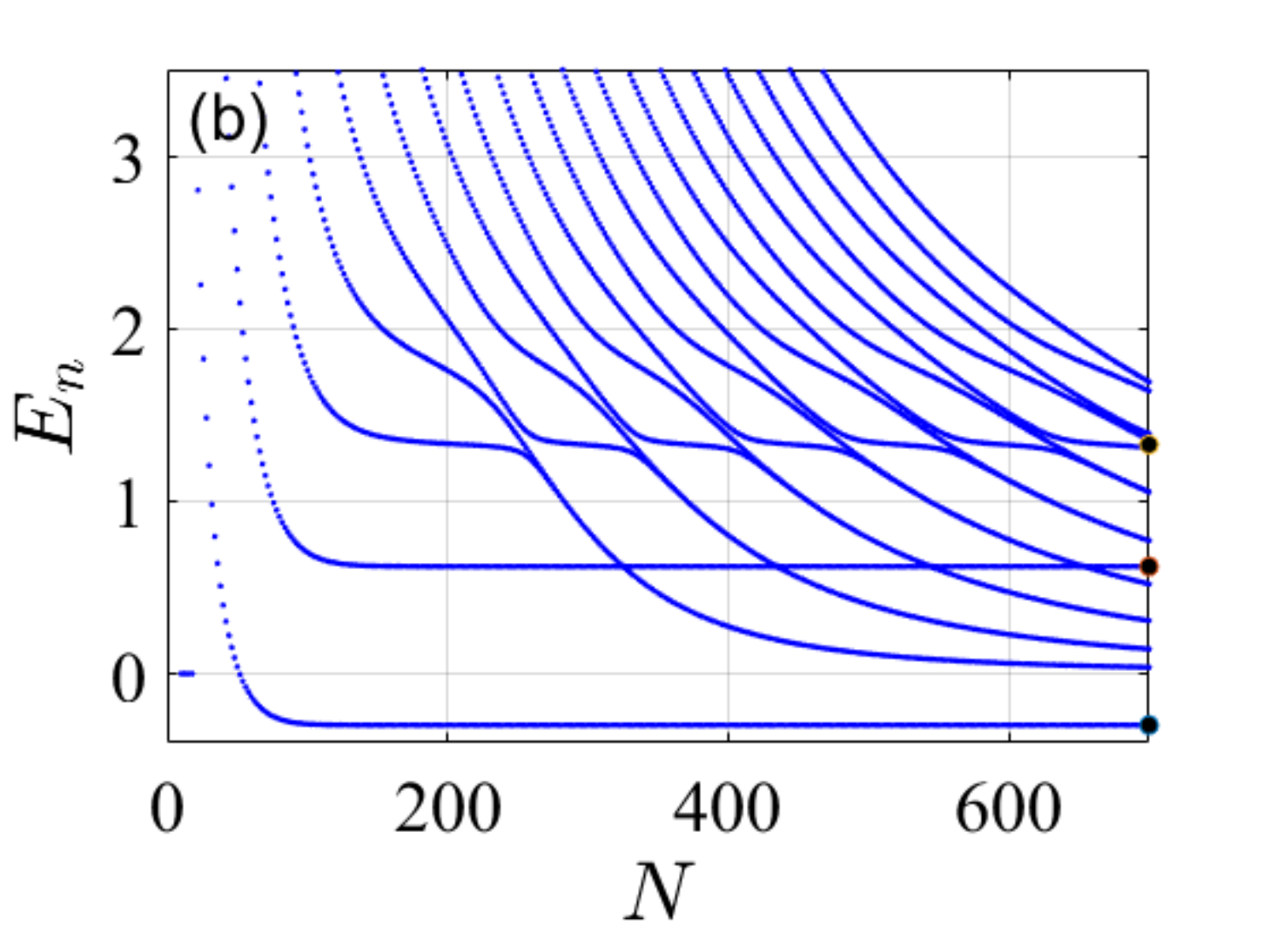}
    \caption{Stabilization graphs for the (a) standard (non-variational) and (b) present (variational) finite difference methods. The plots show the spectrum, $E_n; n=1,2,...$, as a function of the number of grid points $N$ for a model potential $V_r\b x$, Eq. \eqref{eq:V_r}. The full dark points on the right hand side indicate the exact values of the bound and the two nearest resonance energies as calculated by the uniform complex scaling approach. Both methods lead to accurate values as $N$ increases, however, in the standard method, Panel (a), does not allow to distinguish between the resonances and the other quasi-discrete continuum states. In contrast, the variational principle of the present FDM leads to a typical stabilization graph. The stability of the metastable states in the continuum enables identifying them. Their positions (i.e., energies) is determined by the stable region. Model parameters: $\delta x =L_{\text{max}}/N_{\text{max}} \approx 8.5\cdot 10^{-2}$, $L_{\text{max}}=60$ and $N_{\text{max}}=700$. 
    These numerical parameters lead to an error of $10^{-9}$ in the value of bound state, {as well as} $10^{-5}$ and $10^{-2}$ in the values of the two first metastable states, respectively.}
    \label{fig:6}
\end{figure}

{In the following}, the two finite different methods are applied to solve for the spectrum of $H_r=T+V_r$ and the {eigenenergies} are plotted as a function of the number of grid {points} in Fig. \ref{fig:6}. The convergence of the standard {(non-variational)} method is characterized by non-intersecting lines, Fig. \ref{fig:6} Panel (a). As a result, this plot does not {indicate which states are metastable}. In contrast, the present variational FDM produces an informative picture of a typical stabilization graph, Fig. \ref{fig:6}
Panel (b), allowing to distinguish the resonances from the other states in the
quasi-discrete spectrum of the continuum. The possibility to isolate the resonances from the other states in the quasi-continuum can be utilized to calculate the resonance widths. See for example, the calculation of resonance widths (inverse lifetimes) and energies for $L^2$ methods \cite{hazi1970stabilization}, spherical-box quantization\cite{maier1980spherical} and  from the analytical continuation of real stabilization graphs, utilizing a Gaussian basis
\cite{falcetta1991stabilization,falcetta2000ab,falcetta2014assessment,falcetta2016ab,fennimore2016core,landau2016atomic,landau2017ab,bhattacharya2017polyatomic,fennimore2018electronic,gasperich2018strategy,kairalapova2019prediction,thodika2019comparative,landau2019shaping,bhattacharya2019quantum,landau2019clusterization,thodika2020description,chao1990application,lee2020stabilization,bhattacharya2020ab,landau2020ab,carlson2021fresh,slimak2021role,ben2021uniform}.
{This is the primary result of our work, demonstrating that} resonances can be calculated from stabilization graphs obtained by utilizing grid methods and not only by applying basis sets, as previously done.


In order to {obtain} the resonance energies (in an improved accuracy) and the corresponding resonance widths we repeat on the FDM calculations with a uniformly rotated coordinate  in the complex plane. Formally, the procedure maps the $x$ coordinates to  $\{x_i\to x_i\exp(i\theta)\}_{i=1,2,...,N}$, leading to complex eigenvalues. The real part of the eigenvalues that are invariant under the mapping (invariant under a change of $\theta$) are the resonance energies (positions). While the resonance widths are associated with the imaginary parts of the complex eigenvalues multiplied by $-2$ ($E_n\ra \eps_n -i\Gamma/2$). 

For a formal justification for calculating the resonances by a rotation of the coordinated in the complex plane see the text book on non-Hermitian quantum mechanics and the references therein \cite{moiseyev2011non}. Specifically, the works of Aguilar and Combes \cite{aguilar1971class}, and Balslev and Combes \cite{balslev1971spectral}, should be highlighted along with the work of Barry Simon that put the computations of resonances by the complex scaling transformations on a solid mathematical ground \cite{simon1972quadratic,simon1973resonances}.

\begin{figure}[htb!]
    \centering
    \includegraphics[width=6.6cm]{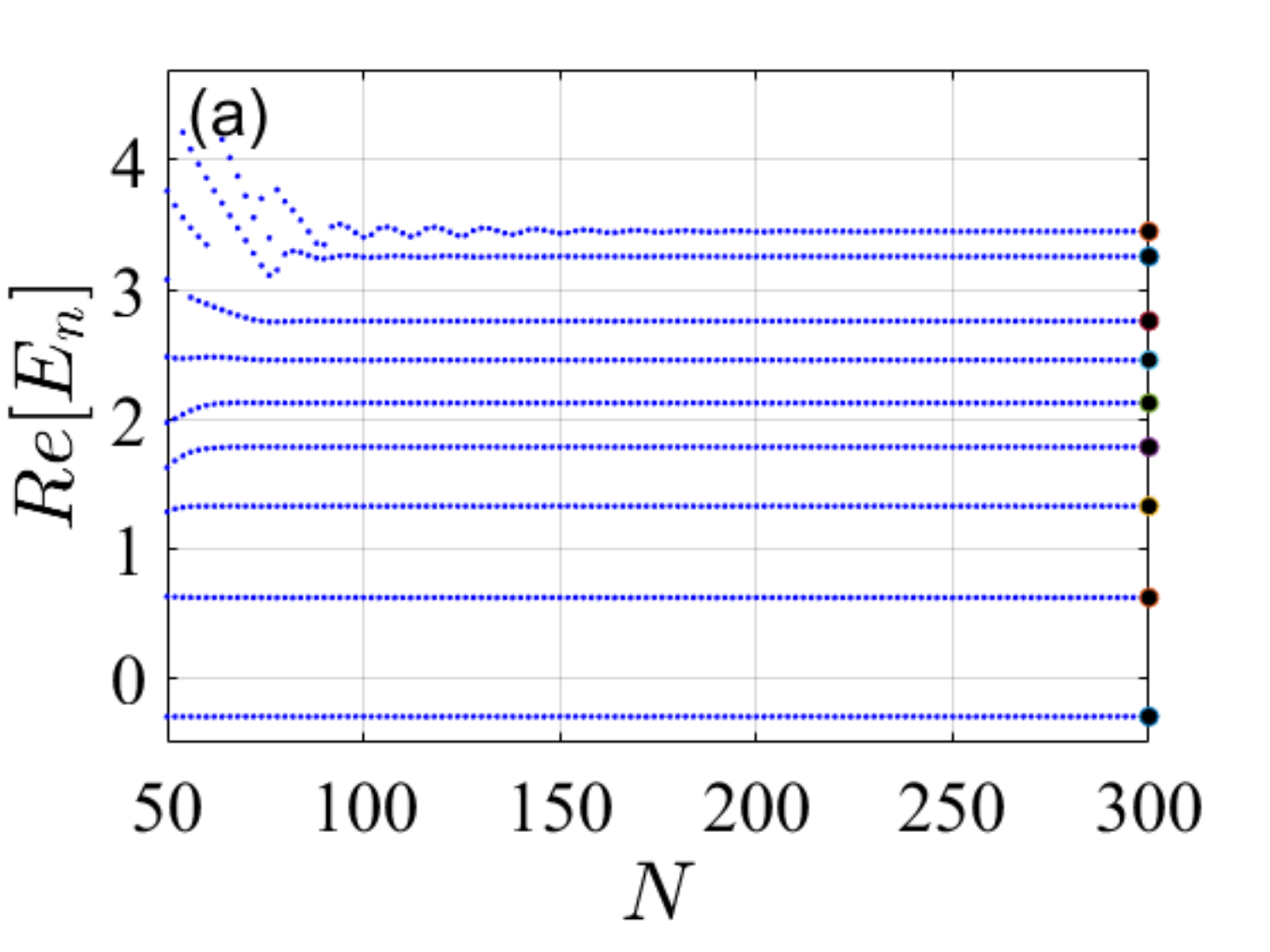}
    \includegraphics[width=6.6cm]{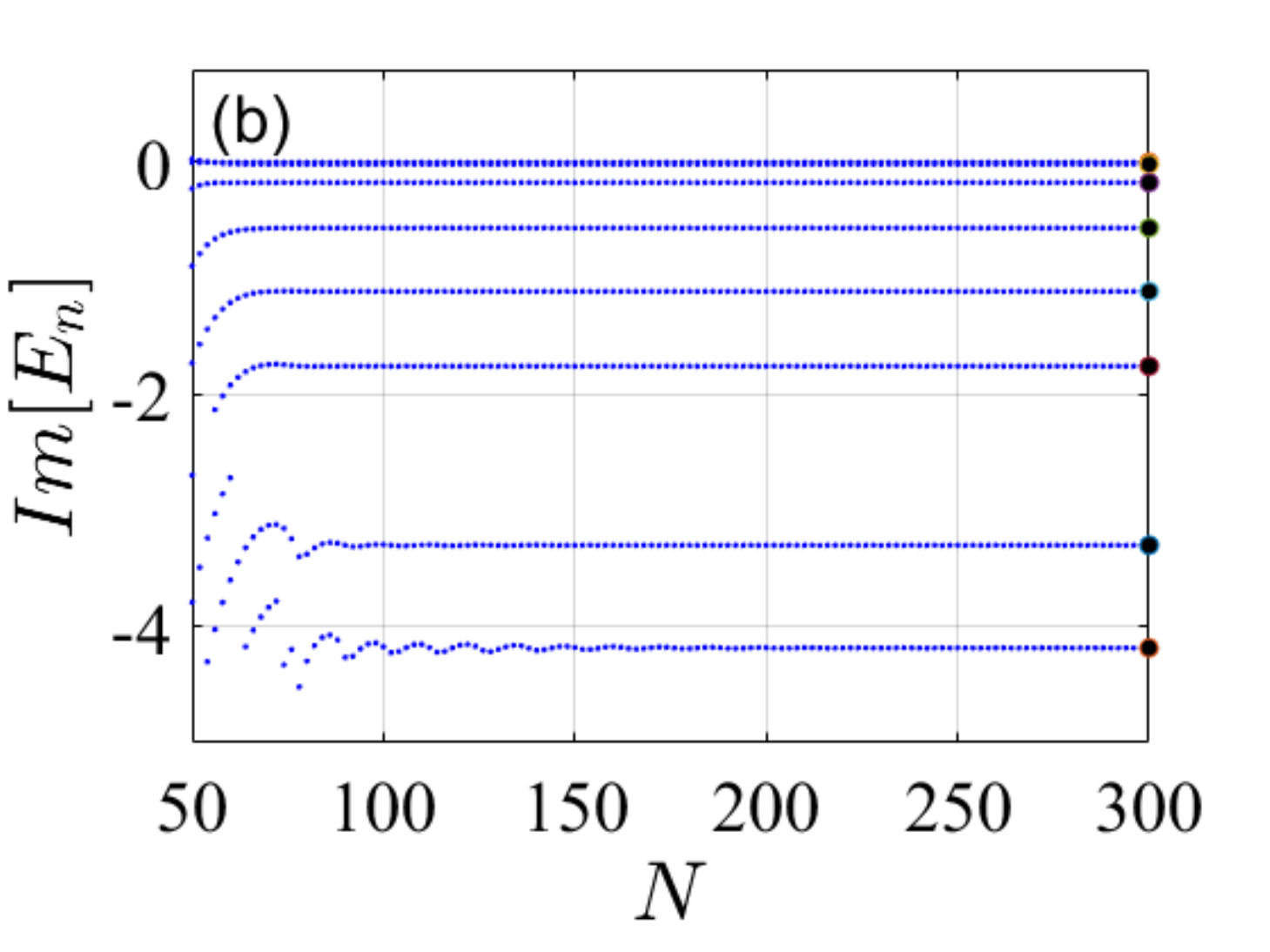}
    \caption{Stabilization graph of the complex scaled spectrum {$\{E_n=\eps_n-i\Gamma_n/2\}$} as obtained by the present FDM. (a) Real part and (b) imaginary part of the spectrum. The grid points were rotated to the complex coordinate plane by an angle $\theta=0.5$, leading to $\{x_i\to x_i\exp(i\theta)\}_{i=1,2,...,N}$.  Exact values of the energies and widths are presented by large black circles on the right hand side. Model parameters: $\delta x = L_{\text{max}}/N_{\text{max}} \approx 0.15$, $L_{\text{max}}=60$ and $N_{\text{max}}=400$.}
    \label{fig:7}
\end{figure}

The results presented in Fig. \ref{fig:7} were obtained by following the complex eigenvalues which their real parts are closest to the resonance values obtained in Fig. \ref{fig:6} ($\theta=0$). This is a crucial property of the present method, as it enables following the convergence of the complex poles as the number of grid points $N$ is increased.  This cannot be achieved by the standard FDM. {The  stability which results from introducing absorbing boundary condition (here obtained by employing complex scaling) has been observed before for the Helium resonances in Ref. \cite{moiseyev1980criteria}, utilizing Hylleraas basis functions.  Therefore, it is expected that similar stabilization graphs will be obtained by grid methods in combination with complex absorbing potentials.}
\section{Concluding remarks}
We show that by fixing the grid spacing in the finite difference method one obtains a variational principle within the finite box approximation. This property allows obtaining a stabilization graphs for the spectrum, which produces an accurate estimation for both the bound ground, excited states as well as the positions of narrow resonances. The stabilization graphs obtained by the presented FDM enable one to distinguish between the metastable (resonance) states that are localized in the interaction region and the other states in the continuum. The later states are not localized and have large amplitudes outside the interaction region. To obtain the resonance width (inverse lifetimes), we  preformed a rotation of the grid points  in the complex plane. This procedure also increases the accuracy of the the resonance positions. 
The calculation accuracy is determined by the grid spacing. As the grid spacing decreases the accuracy increases at the expense of increasing the number of grid points required for convergence.

{Overall,} the present study demonstrates how a simple change in a known numerical method (FDM in our case) might increase the broadness of its application. Our results suggest that grid methods should be {added to the Gaussian basis set} in the electronic structure quantum chemistry packages.  {This will allow introducing complex absorbing boundary conditions in the non-interacting region of the Hamiltonian. This approach paves} the way for the calculations of the complex poles of the scattering matrix. Such poles can be associated with the peaks in measured reaction rates in cold molecular collisions.

\acknowledgments{We thank Ronnie Kosloff for fruitful discussions. This research was supported by the Adams Fellowship  Program of the Israel Academy of Sciences and Humanities. The work is based on a project given in a course on Numerical Methods in Quantum Mechanics given at the Technion by ZOOM during the Coronavirus lockdown. NM  congratulate Frank on his 80 birthday and  wish him many more healthy years and to continue enjoying doing good science as he does. {The presented results are based on Matlab codes, which will be gladly sent upon request.}}

\appendix
\section{{Derivation of the representation of the Laplacian by the standard finite difference method}}
\label{appendix}
{The kinetic energy operator $T$ is presented by a $N\times N$ matrix, which employs $m$ (where $m=3,5,..$) grid points to approximate the representation of the Laplacian operator at each point. This approximation results in a discretized spectrum of the kinetic energy operator, given by  $E_n=c\b{n/L}^2$, where  $n\in \mathbb{N}$ is the quantum number}, $L=x_N-x_1$ is the size of the box which discretizes the kinetic energy spectrum and the proportionality constant $c$ is problem dependent. In the solution of the time-independent Schr\"odinger equation the proportionality constant equals to $\pi^2\hbar^2/(2\mu)$. Commonly, in a grid representation, the matrix representing the potential energy is diagonal with values $V\b{x_i}$, where $x_{i}$  denotes the coordinate of the $i^{\mathrm{th}}$ nodal point.

For the sake of clarity we give below a short description of the derivation of the {approximate matrix representation of the Laplacian operator}. Consider a 1–dimensional evenly spaced grid made up of $N$ nodal points with a total length $L$. The spacing between adjacent nodal points is given by
$\Delta x=\frac{L}{N-1}$. We wish to approximate the second order derivative of the wave function $\psi\b{x}$ at $x=x_{i}$, compactly denoted as $\psi_i$. For this purpose, we write the truncated Taylor series expansion around $x_{i}$ using $j$ nodal steps, explicitly written as
\begin{equation}
    \psi_{i+j}=\sum_{k=0}^{m-1}\frac{\b{j\Delta x}^{k}}{k!}\left.\frac{d^{k}\psi}{dx^{k}}\right|_{x_{i}}~~,
    \label{eq:2}
\end{equation}
where $j\in \sb{-l,l}$ {and $m=2l+1$}. This results in a linear system of equations which relates the vector of the nodal values of the function $\v \psi$ and the vector of its derivatives $\v{\psi}^{\left(D\right)}$
\begin{equation}
    \v{\psi}=\left\{ \psi_{i+j}\right\} _{j=-l}^{l}~~,~~\v{\psi}^{\left(D\right)}=\left\{ \left(\Delta x\right)^{k} \left.\frac{d^{k}\psi}{dx^{k}}\right|_{x_{i}}\right\}_{k=0}^{m-1}~~.
\end{equation}
{Notice that the first and last elements of $\v{\psi}$ are $\psi_{i-l}$ and  $\psi_{i+l}$, respectively}. The nodal values and its derivatives are related through the matrix $A^{m\times m}$, with elements $A_{j+l+1,k+1}={j^{k}}/{k!}$:
\begin{equation}
    \v{\psi}=A\v{\psi}^{\left(D\right)}~~.
    \label{eq:4}
\end{equation}
By inverting Eq. \eqref{eq:4}, we isolate
the second order derivative, which is proportionate to the third element of $\v{\psi}^{\b{D}}$. This leads to  a linear combination of the nodal values, with weights $w_{j}=\mathrm{A}_{3,j+l+1}^{-1}$.
The derivative is then explicitly written as
\begin{equation}
     \left(\Delta x\right)^{2} \left.\frac{d^{2}\psi}{dx^{2}}\right|_{x_{i}}=\sum_{j=-l}^{l}w_j\psi_{i+j}~~.
\end{equation}

\section{{Convergence of the kinetic energy representation by the standard finite difference method}}
\label{apsec:plot}
\begin{figure}[htb!]
    \centering
    \includegraphics[width=8cm]{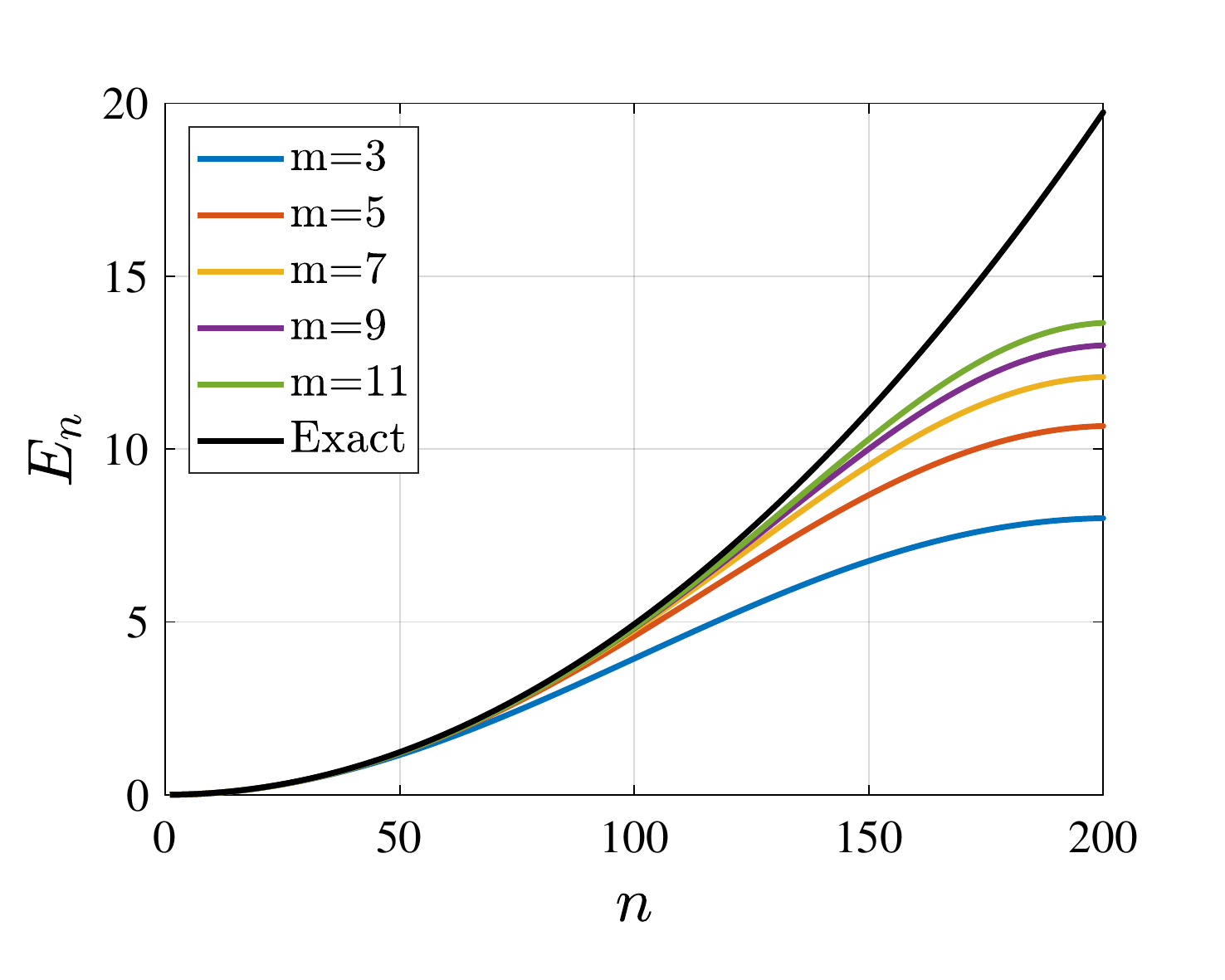}
    \caption{{Eigenvalues of the kinetic energy operator $E_n$, $n=1,2,...$, as a function of the quantum number $n$ as obtained by the standard FDM. The different plots correspond to different values of grid points employed to calculate  the second order derivative, $m$ (Eq. \eqref{eq:2}), while keeping the total number of grid points and the box size constant. As $m$ increases the accuracy of the second order derivative is increased. The chosen parameters are: $\hbar = 1$, $\mu =1 $, $N=201$ and $L=100$. Note that this figure is given for illustration reasons and is not novel (see for example Fig. 1 in Ref. \cite{maragakis2001variational}).}}
    \label{fig:KE_m}
\end{figure}

{The eigenvalues of $T$ converge towards the analytical result with an increase in the number of grid points included in the calculation $m$. The different matrices $T\b m$ are diagonalized for a constant number of grid points $N$, to obtain the eigenenergies $E_n$ for different values of $m$.  Figure \ref{fig:KE_m} presents the eigenvalues of the matrices $T\b{m}$ in increasing order. As the $m$ increases the representation of the kinetic energy operator becomes more accurate and the corresponding eigenenergy curves converge to the exact result. Therefore, the value of $m$ has a major influence on the derivation of the upper  bound within a specific potential.}

\break

\end{document}